\documentclass[journal]{IEEEtran}
\usepackage{cite}

\ifCLASSINFOpdf
   \usepackage[pdftex]{graphicx}
\else
\fi
\usepackage{amsmath}
\usepackage{array}
\begin{document}
%
\title{Flexible Coherent Optical Access: Architectures, Algorithms, and Demonstrations}
%
%
%

\author{Ji Zhou,  
        Zhenping Xing,
        Haide Wang,
        Kuo Zhang,
        Xi Chen,
        Qiguang Feng,
        Keshuang Zheng,
        Yijia Zhao,
        Zhen Dong,
        Tao Gui,
        Zhicheng Ye,
        and Liangchuan Li

\thanks{Manuscript received xxxx; revised xxxx. (Corresponding authors: Liangchuan Li).}
\thanks{J. Zhou and H. Wang are with the Department of Electronic Engineering, College of Information Science and Technology, Jinan University, Guangzhou 510632, China.}
\thanks{Z. Xing, K. Zhang, X. Chen, Q. Feng, K. Zheng, Y. Zhao, Z. Dong, T. Gui, Z. Ye, and L. Li are with Optical Research Department, Huawei Technologies Co Ltd, Dongguan, 523808, China.}
}

\maketitle

\begin{abstract}
To cope with the explosive bandwidth demand, significant progress has been made in the ITU-T standardization sector to define a higher-speed passive optical network (PON) with a 50Gb/s line rate. Recently, 50G PON becomes mature gradually, which means it is time to discuss beyond 50G PON. For ensuring an acceptable optical power budget, beyond 50G PON will potentially use coherent technologies, which can simultaneously promote the applications of flexible multiple access such as time/frequency-domain multiple access (TFDMA). In this paper, we will introduce the architectures, algorithms, and demonstrations for TFDMA-based coherent PON. The system architectures based on an ultra-simple coherent transceiver and specific signal spectra are designed to greatly reduce the cost of ONUs. Meanwhile, fast and low-complexity digital signal processing (DSP) algorithms are proposed for dealing with upstream and downstream signals. Based on the architectures and algorithms, we experimentally demonstrate the first real-time TFDMA-based coherent PON, which can support at most 256 end users, and peak line rates of 100Gb/s and 200Gb/s in the upstream and downstream scenarios, respectively. In conclusion, the proposed technologies for the coherent PON make it more possible to be applied in the future beyond 50G PON.
\end{abstract}


%
\IEEEpeerreviewmaketitle

\section{Introduction}
\begin{figure}[!t]
\centering
\includegraphics[width=\linewidth]{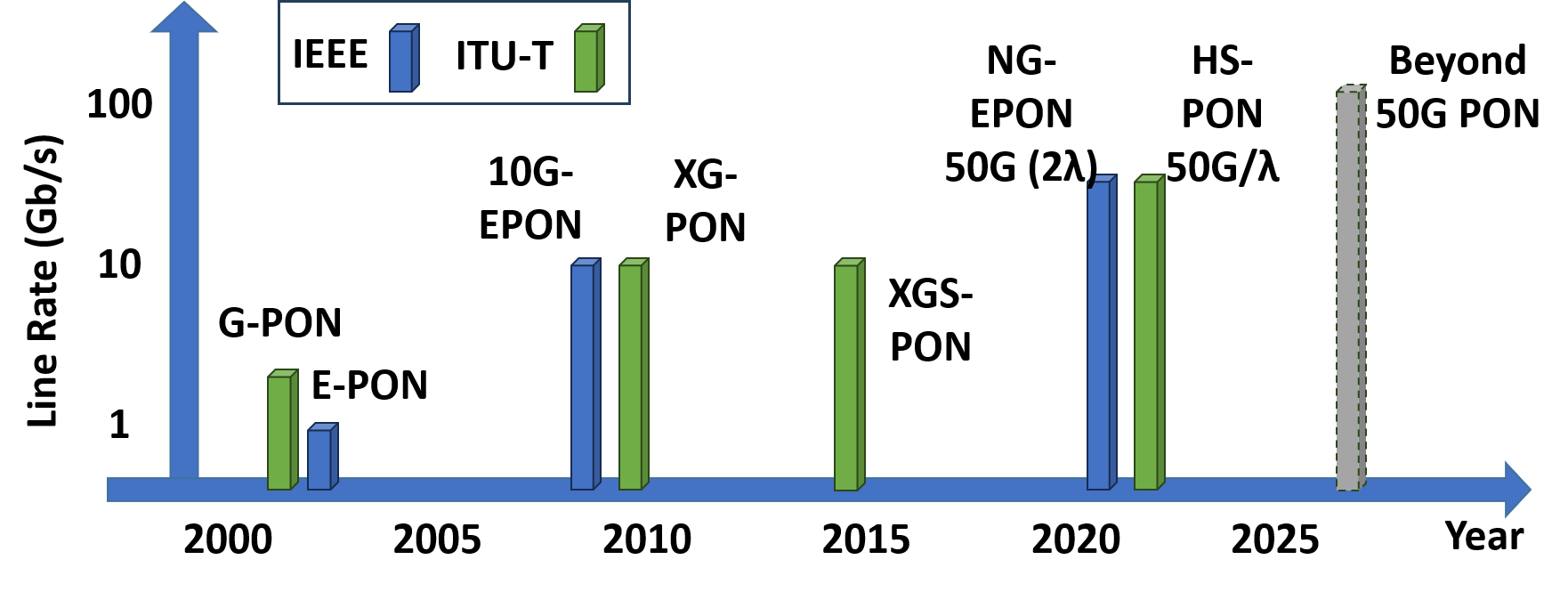}
\caption{The roadmap for IEEE and ITU-T standards of PONs.}
\label{Roadmap}
\end{figure}

\IEEEPARstart{D}{riven} by the sustainably growing traffic demand, the line rate of passive optical network (PON) is rising steadily. Fig. \ref{Roadmap} shows the roadmap for IEEE and ITU-T standards of PON. Significant progress has been made in the ITU-T standardization sector to define higher-speed (HS) PON with a line rate of 50Gb/s \cite{ITUT202150gigabit,9123509}. Recently, 50G PON  becomes mature gradually, which means it is time to discuss beyond 50G PON \cite{ITUT2023b50gigabit, zhang2022coherent, suzuki2022digital}. For ensuring an acceptable optical power budget, beyond 50G PON will potentially use coherent technologies \cite{teixeira2016coherent, lavery2018recent, faruk2020coherent}. Among all the potential solutions, a coherent PON based on flexible time-domain/frequency-domain multiple access (TFDMA) is one of the most appealing options, which can combine the statistical multiplexing capability of TDMA and the dedicated frequency allocation capability of FDMA \cite{zhang2021efficient, xu2022intelligent, zhang2023low}. Based on digital subcarrier multiplexing (DSCM), FDMA-based coherent PON allows each optical network unit (ONU) to transmit and receive only a subset of subcarriers, which significantly reduces the bandwidth of its transceiver\cite{welch2022digital, welch2022digital-1, hosseini2022multi}. Other benefits of coherent PON include: 1) it improves the receiver sensitivity due to the use of a local oscillator (LO), and 2) it can use C-band wavelength resources that have not been used in previous PONs since the dispersion can be effectively compensated by digital signal processing (DSP) \cite{xing2022first, wei2022time, zhang2023flexible}.

Unfortunately, the use of a full coherent transceiver at each ONU is still cost-prohibitive for the PON scenario. To further reduce the cost of an ONU, it has been proposed to use a single-polarization heterodyne receiver based on Alamouti coding rather than a fully coherent receiver in the downstream scenario \cite{faruk2022experimental, hraghi2022analysis, li2022bidirectional}. In the upstream scenario, The ONU can also use a single Mach-Zehnder modulator (MZM) instead of a dual-polarization in-phase and quadrature MZM. Such an ultra-simple coherent transceiver contains one digital-to-analog converter (DAC), one single MZM, one optical coupler, one balanced photo-detector (BPD), one analog-to-digital converter (ADC), and two lasers. Two lasers become the major cost in the ultra-simple coherent transceiver \cite{gaudino2012use}. The high-cost external cavity laser used in traditional coherent transceivers cannot meet the requirement of the ultra-simple coherent transceiver.

In this paper, we experimentally demonstrate the first real-time TFDMA-based coherent PON using an ultra-simple transceiver. The proposed PON can support a splitting ratio up to 1:256, and peak line rates of 100Gb/s and 200Gb/s in the upstream and downstream scenarios, respectively. In addition, we prove that high-precision DSP-aided frequency locking makes the cost-effective distributed feedback (DFB) laser feasible for the ultra-simple coherent transceiver. This paper is an extended version of our published post-deadline paper in OFC 2023 \cite{Xing:23}. More detailed information about the DSP algorithms is added in this extended version.

The main contributions of this paper are as follows: 
\begin{itemize}
	\item The system architectures based on an ultra-simple coherent transceiver and specific signal spectra are designed to greatly reduce the cost of ONUs.
	\item  We propose fast and low-complexity DSP algorithms for effectively processing upstream and downstream signals in TFDMA-based coherent PON.
    \item We demonstrate the first real-time TFDMA-based coherent PON, which can support at most 256 end users, and peak line rates of 100Gb/s and 200Gb/s in the upstream and downstream scenarios, respectively.
\end{itemize}

The remainder of this paper is organized as follows. The system architectures and specific signal spectra are shown in Section \ref{SectionII}. In Section \ref{SectionIII}, the DSP algorithms for TFDMA-based coherent PON are introduced in detail.  In Section \ref{SectionIV}, the experimental setups and results are given. Finally, the paper is concluded in Section \ref{SectionV}.

\begin{figure}[!t]
\centering
\includegraphics[width=\linewidth]{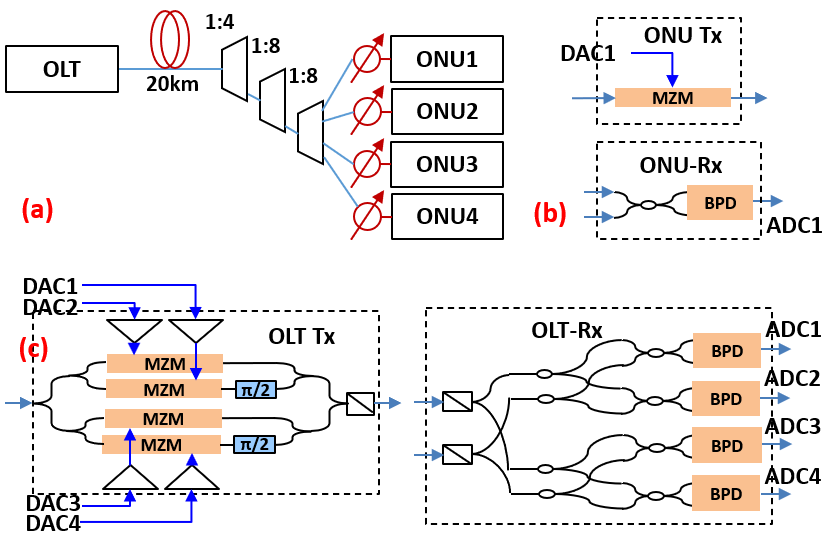}
\caption{(a) The architecture diagram of the coherent PON with 256 splitter ratio. (b) Transmitter (Tx) and receiver (Rx) devices of the ultra-simple coherent transceiver at the ONU. (c) Tx and Rx devices of the full coherent transceiver at the OLT.}
\label{Experiment}
\end{figure}

\section{System Architecture and Spectra Design} \label{SectionII}
In this section, we will introduce the system architectures with an ultra-simple coherent transceiver at the ONU and a full coherent transceiver at the OLT. Meanwhile, the specific signal spectra are designed for the upstream and downstream scenarios in the TFDMA-based coherent PON.

The system architecture of the coherent PON is shown in Fig. \ref{Experiment}(a), which is designed to support at most 256 ONUs by using 1:4, 1:8, and 1:8 passive optical splitters. Without loss of generality, we test 4 ONUs on the edge of the network. The transmitter and receiver devices of the ultra-simple coherent transceiver at the ONU are depicted in Fig. \ref{Experiment}(b). The ultra-simple coherent transceiver consists of one DAC, one single MZM, one optical coupler, one ADC, one single BPD, and two DFB lasers. Two DFB lasers are used as the LO and optical carrier for the downstream and upstream scenarios, respectively. Therefore, Single-polarization heterodyne detection and unidimensional signal generation are implemented by the ultra-simple coherent transceiver. For the downstream scenario, an Alamouti-coding signal should be received at the ONU to avoid the state-of-polarization (SOP)-caused signal disappearance. For the upstream scenario, the transmitted signal at the ONU should be a real-valued carrier-less amplitude phase (CAP) signal to tolerate direct-current (DC) leakage. For generating the Alamouti-coding signal and detecting the CAP signal at the OLT, a full coherent transceiver can be deployed, as shown in Fig. \ref{Experiment} (c). The full coherent transceiver at the OLT makes it possible to gradually evolve the line rate by updating the transceiver at the ONU.

\begin{figure}[!t]
\centering
\includegraphics[width=\linewidth]{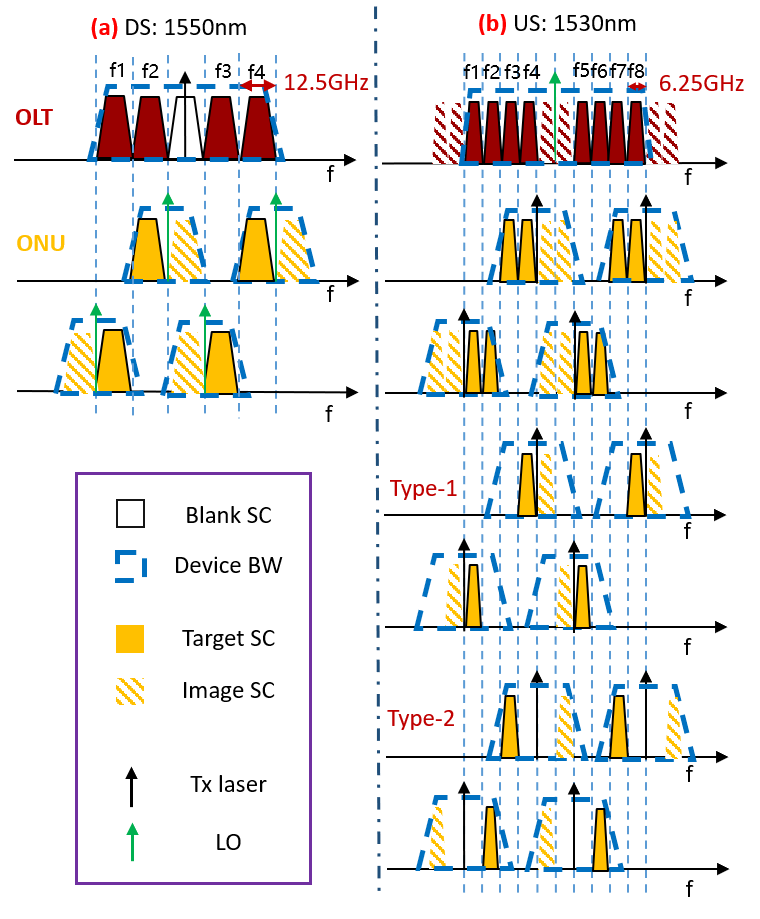}
\caption{(a) Designed spectra with a bandwidth granularity of 12.5GHz for the downstream scenario. (b) Designed spectra with a bandwidth granularity of 6.25GHz for the upstream scenario. SC: Subcarrier. BW: Bandwidth. DS: Downstream. US: Upstream.}
\label{Spectrum}
\end{figure}

Figure \ref{Spectrum} (a) shows the designed signal spectra with a bandwidth granularity of 12.5GHz for the downstream scenario. In the downstream scenario, 5$\times$12.5Gbaud digital subcarriers with a granularity of 12.5GHz are filled in the whole bandwidth. Only 4 digital subcarriers carry data, and the central subcarrier is blank for implementing the heterodyne detection. Then, Alamouti coding with a rate of 1/2 is used to implement the single-polarization detection. However, the capacity of the Alamouti-based dual-polarization optical system is equivalent to that of a single-polarization optical system with the same bandwidth. Therefore, the peak line rate of the downstream scenario is only 200Gb/s when 16QAM is modulated on 50GHz bandwidth. We use probabilistic constellation shaping 16QAM (PCS-16QAM) to achieve flexible-rate adaption from 100Gb/s to 200Gb/s for making full use of optical power budget \cite{liu2023flexible, zhou2022100g,shen2023demonstration}

In the upstream scenario, each ONU transmits a real-valued CAP signal, as shown in Fig. \ref{Spectrum} (b). The reasons are that 1) a single MZM can only modulate a real-valued signal, and 2) a guard band around DC frequency is required. The MZM at the ONU is biased at the null point. To achieve steep rising and falling edges, the burst-mode signals are generated by switching on/off the DAC rather than the laser. However, due to the limited extinction ratio of the MZM, the DC leakage of laser power is not negligible when the DAC is off, hence the guard band is necessary. We further break each 12.5Gbaud subcarrier into 2$\times$6.25Gbaud subcarriers for providing fine-granularity transmission in the upstream scenario. More frequency resources can be provided for dedicated usage in the upstream scenario. Dedicated subcarriers allow high-end users to get free from rogue ONUs, which only exist in the upstream scenario. Each ONU can transmit either the inner (Type-1) or the outer (Type-2) subcarrier in CAP with two subcarriers. In our work, only quadrature phase shift keying (QPSK) was modulated for the upstream scenario to obtain a peak line rate of 25Gb/s per ONU and a total peak line rate of 100 Gb/s from the OLT’s perspective.

In conventional FDMA-based PON, four subcarriers for the downstream scenario and eight subcarriers for the upstream scenario cannot support the bandwidth allocation for 256 ONUs. However, if the subcarrier number is increased, it is hard to compensate for the phase noise by the DSP algorithm, and the guard band increases to decrease the spectral efficiency. The TDMA is added to the subcarriers to implement TFDMA, which is a feasible method for increasing the number of ONUs. In the TFDMA, the subcarriers can be individually allocated to low-latency and bandwidth-hungry ONUs. Meanwhile, for the common ONUs, the subcarrier can be divided into some time slots to provide flexible bandwidth allocation. In conclusion, the system architectures, specific signal spectra, and TFDMA can be used to support low-latency or low-cost ONUs.

\section{DSP Algorithms for Coherent PON} \label{SectionIII}
Traditional coherent DSP algorithms are not suitable for TFDMA-based coherent PON. In this section, we will introduce the specially designed DSP algorithms for the TFDMA-based coherent PON. The DSP algorithms for the upstream and downstream scenarios work in the burst mode and continuous mode \cite{wang2023fast, savory2010digital, zhang2020efficient}, respectively. For the upstream scenario, burst-mode DSP algorithms based on training sequences are used to achieve fast convergence for reducing the overhead and improving spectral efficiency. For the downstream scenario, the continuous-mode DSP algorithms at the ONU are sensitive to computational complexity and power consumption. Only a part of the DSP algorithms should be always turned on to track the dynamical distortions, such as frequency-offset estimation (FOE), timing error detection, coefficient estimation of the equalizer, and carrier-phase estimation (CPE). At the time slots of other ONUs, the DSP algorithms processing the payloads and forward error correction (FEC) can be turned off to reduce power consumption. In addition, frame synchronization is required only once when the ONU is initially registered.

\subsection{Frame Detection and Coarse FOE} \label{SectionII-A}
In this subsection, periodic sequences are designed to simultaneously implement frame detection and coarse FOE.
\subsubsection{Frame Detection}
For the upstream scenario, frame detection is required to recognize two symmetric frequency tones for confirming the arrival of burst fame. The symmetric frequency tones are generated by a periodic sequence. Frame detection is not necessary for the continuous-mode downstream scenario. The frame detection can be implemented by the following steps. Firstly, a sliding window operation is applied to the received signal. Then, the extracted signal is down-sampled and transferred to the frequency domain by a fast Fourier transform (FFT). Finally, the average power of the non-zero points is calculated. The frequency points with a power less than average power are filtered out. These mentioned operations are repeated three times to find the accurate frequency tones. 

\subsubsection{Coarse FOE}
\begin{figure}[!t]
\centering
\includegraphics[width=\linewidth]{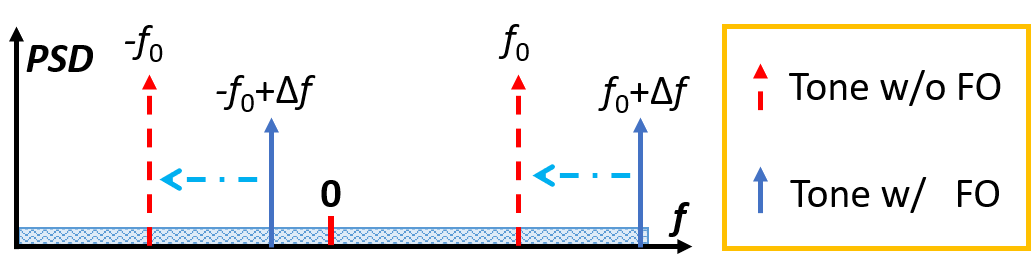}
\caption{The frequency tones with and without frequency offset (FO).}
\label{FOE}
\end{figure}

After the frame detection, the frequency offset should be compensated to ensure the signal spectrum within the frequency-domain range of the matched filter. Fig. \ref{FOE} shows the frequency tones with and without frequency offset. The detected frequency tones after the frame detection can be used to estimate the frequency offset by
\begin{equation}
\Delta f_{\text {Coarse}}=\frac{1}{2} \times\left(f_{1}+f_{2}\right)
\end{equation}
where $f_{1}$ and $f_{2}$ are the frequencies of the $-f_0+\Delta f$ and $f_0+\Delta f$, respectively. $\Delta f$ is the actual frequency offset. The FFT size confines the accuracy of the FOE, which is limited by the parallelism of real-time implementation. For example, when the FFT size is 32 and the signal bandwidth is 8GHz, the resolution of one frequency point is only 250MHz, which may lead to $\pm125$MHz deviation for the FOE. Therefore, the FOE based on the detected frequency tones is coarse, and more fine FOE is required.

\subsection{Burst- and Continuous-Mode Timing Recovery}
For the upstream scenario of the TFDMA-based coherent PON,  burst-mode timing recovery should be used to accelerate convergence to reduce the overhead. For the downstream scenario, the timing recovery works in continuous mode.

\begin{figure}[!t]
\centering
\includegraphics[width=\linewidth]{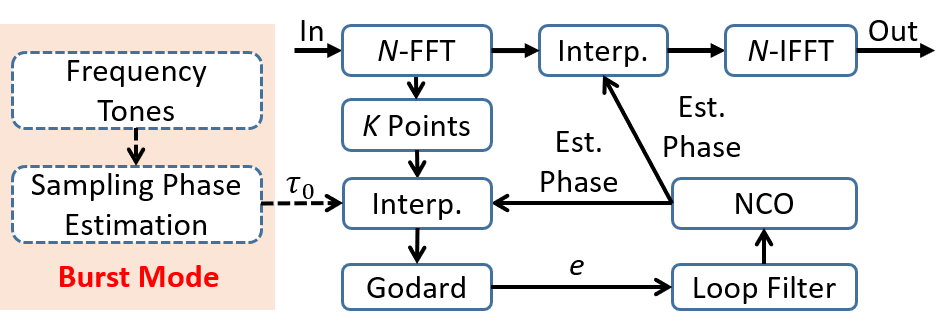}
\caption{The burst-mode timing recovery with sampling phase initialization for the upstream scenario, and continuous-mode timing recovery for the downstream scenario. Interp.: Interpolation. NCO: Numerically controlled oscillator. Est.: estimation.}
\label{TR}
\end{figure}

Fig. \ref{TR} shows the burst-mode timing recovery with sampling phase initialization for the upstream scenario, and continuous-mode timing recovery for the downstream scenario.
The common structure of burst-mode and continuous-mode timing recovery is the frequency-domain Godard algorithm. For the upstream scenario, an appropriate sampling phase initialization plays a significant role in reducing the convergence time of timing recovery. Fortunately, the initialized sampling phase offset $\tau_0$ can be estimated by using two frequency tones in Subsection \ref{SectionII-A}, which can be calculated by
\begin{equation}
\tau_0 = \frac{1}{4\pi f_0}\text{arg}[R(f_0)\times R^*(-f_0)]
\end{equation}
where $(.)^*$ denotes the conjugate operation. $R(f)$ is the received frequency tones with the initialized sampling phase offset such as $\delta(f-f_0)e^{j2\pi f_0 \tau+\phi_0}$ and $\delta(f+f_0)e^{-j2\pi f_0 \tau+\phi_0}$. $\phi_0$ is the phase noise, which does not influence the estimation of sampling phase offset. The initialized sampling phase offset is injected into the frequency-domain Godard algorithm to reduce the convergence time. 

The frequency-domain Godard algorithm is implemented using the signal spectrum $\mathbf{S}$ after match filtering with a roll-off factor $\beta$. The timing error is estimated as
\begin{equation}
e=\sum_{k=\frac{(1-\beta)}{2 sps} N}^{\frac{(1+\beta)}{2 sps} N-1} \operatorname{Im}\left[S_k \cdot S_{k+(1-1/sps) N}^*\right]
\label{fractional_TR}
\end{equation}
where $\operatorname{Im}(\cdot)$ denotes the imaginary part of a complex value. $sps$ is the samples per symbol. $N$ is the number of the frequency points after the $N$-FFT, which should choose the value that allows the upper and lower bounds to be integers. Based on Eq. (\ref{fractional_TR}), only $K$ frequency points are used to estimate timing error with relatively low complexity where $K$ is equal to $\beta N/sps-1$. Finally, the estimated phase is updated after every iteration.

\subsection{Frame Synchronization and Fine FOE}
In this subsection, training sequences are designed to simultaneously implement frame synchronization and fine FOE.
\subsubsection{Frame Synchronization} 
Frame synchronization is implemented based on the specially designed sequences $[\mathbf{s_1}~\mathbf{s_2}]$ and $[\mathbf{s_3}~\mathbf{s_4}]$ for X polarization and Y polarization, respectively. The $\mathbf{s_1}$ and $\mathbf{s_3}$ consist of QPSK symbols with different pseudorandom binary sequences. The $\mathbf{s_2}$ and $\mathbf{s_4}$ are separately generated by
\begin{equation}
s_{2/4}(i) = s_{1/3}(i) \times pn(i)
\end{equation}
where $i$ is from 1 to $L$. $\mathbf{pn}$ is a pseudo-random noise (PN) sequence with the length of $L$. Finally, the specially designed sequences repeat three times to generate the successive sequences $\left[\mathbf{s_{1/3}}~\mathbf{s_{2/4}},~\mathbf{s_{1/3}}~\mathbf{s_{2/4}},~\mathbf{s_{1/3}}~\mathbf{s_{2/4}}\right]$.

Frame synchronization is realized by using a sliding window and the cross-correlation on each polarization, which is expressed as
\begin{equation}
P(d) = \sum_{i = 0}^{L-1}r(d + i) \times \left[pn(i)\times r^{*}(d + i + L)\right]
\end{equation}
where $\mathbf{r}$ is the received signal of $\mathbf{s}$. The frame synchronization is based on the timing metric, which is calculated by 
\begin{equation}
M(d) = \frac{|P(d)|^2}{P_r^2(d)}.
\end{equation}
where the half-symbol energy $P_r(d)$ is defined as
\begin{equation}
P_r(d) = \frac{1}{2}\sum_{i = 0}^{2L-1}|r(d + i)|^2.
\end{equation}

There are five sharp peaks in the $M(d)$. To enhance the tolerance for noise, $M(d)$ with $0$, $L$, $2L$, $3L$ and $4L$ delay are stacked over, which can be expressed as
\begin{equation}
M'(d) = \sum_{i = 0}^{4}M(d + i \times L),
\end{equation}
The highest-peak position of $M'(d)$ is the accurate frame synchronization position. 

\subsubsection{Fine FOE} 
The successive sequences for frame synchronization can also be used to implement fine FOE. For convenient analysis, we define $\left[\mathbf{s_{1/3}}~\mathbf{s_{2/4}},~\mathbf{s_{1/3}}~\mathbf{s_{2/4}},~\mathbf{s_{1/3}}~\mathbf{s_{2/4}}\right]$ as [$\mathbf{s_{B1}}$, $\mathbf{s_{B2}}$, $\mathbf{s_{B3}}$]. The received signal of $\mathbf{s_{B1}}$, $\mathbf{s_{B2}}$ and $\mathbf{s_{B3}}$ can be defined as $\mathbf{r_{B1}}$, $\mathbf{r_{B2}}$ and $\mathbf{r_{B3}}$, respectively. When the fine frequency offset is considered, the received signal can be defined as
\begin{equation}
r_{\text{B}i}(n) = s_{\text{B}i}(n)\cdot\exp({j\frac{2\pi n}{R_s} \cdot \Delta f_{\text{Fine}}})
\end{equation}
where $i$ is from 1 to 3. $\Delta f_{\text{Fine}}$ is the fine frequency offset. $R_s$ is the baud rate. The fine FOE estimates the frequency offset by \cite{morelli1999improved}
\begin{equation}
\begin{aligned}
\Delta f_{\text{Fine}} &= \frac{R_s}{4\pi L}\cdot\arg{\left(\mathbf{r_{B2}} \times \mathbf{r_{B1}^H} + \mathbf{r_{B3}} \times \mathbf{r_{B2}^H}\right)}\\
&=\frac{R_s}{4\pi L}\cdot\arg{\left[2|\mathbf{s_{B1}}|^2\cdot\exp\left({j\frac{4\pi L}{R_s}\cdot \Delta f_{\text{Fine}}}\right)\right]}
\end{aligned}
\end{equation}
where $\arg(\cdot)$ represents the operation of taking the angle of a complex value and $(\cdot)^\mathbf{H}$ denotes the conjugate transpose operation. Due to the average effect, the fine FOE is more accurate with the increase of $L$. However, the range of fine FOE is from $-R_s/4L$ to $R_s/4L$, which decreases with the increase of $L$. Therefore, the $L$ should be set to a suitable value for balancing the accuracy and range. When the $L$ is set to $10$ and $R_s$ is set to 8Gbaud, the range of fine FOE is from $-200$MHz to $200$MHz. If the frequency offset of the laser is locked within $\pm 100$MHz, the coarse FOE is not required.

\begin{figure}[!t]
\centering
\includegraphics[width=\linewidth]{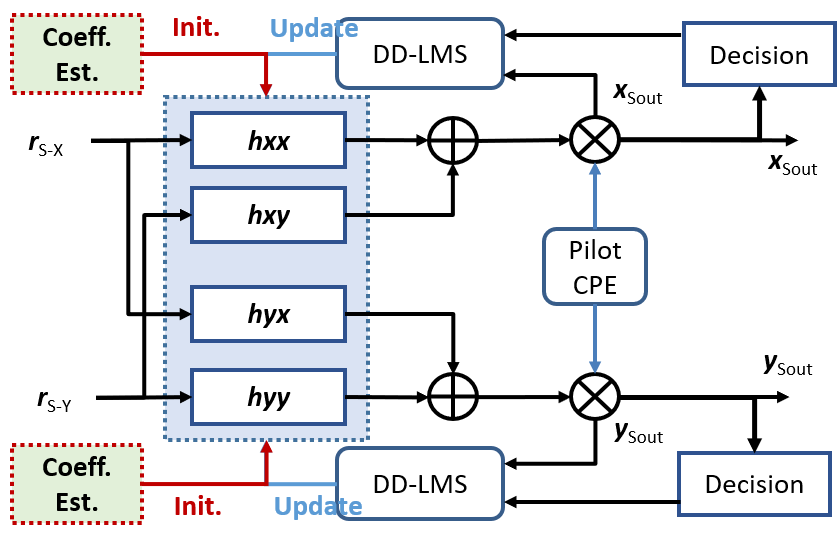}
\caption{The block diagram for the burst- and continuous-mode multiple-input-multiple-output (MIMO) equalizer.}
\label{MIMO}
\end{figure}

\subsection{Burst- and Continuous-Mode MIMO Equalizer}
For the upstream scenario, the burst-mode multiple-input-multiple-output (MIMO) equalizer is used to accelerate convergence for reducing the overhead. For the downstream scenario, the MIMO equalizer works in continuous mode. Fig. \ref{MIMO} shows the block diagram for the burst- and continuous-mode MIMO equalizer. Different from the continuous-mode MIMO equalizer, the burst-mode MIMO equalizer has a coefficient initialization. For fast converging the coefficients, LS-based coefficient estimation is implemented using a designed const amplitude zero auto-correlation training sequence \cite{pittala2014training}. After the coefficient initialization, it switches to decision-directed-LMS (DD-LMS) algorithm to track the coefficients. For the continuous-mode MIMO equalizer, the DD-LMS algorithm tracks the coefficients all the time. Before the DD-LMS algorithm, the carrier-phase noise should be estimated and compensated based on a pilot-based CPE. In our work, the MIMO equalizer at the ONU should be specially designed for dealing with Alamouti-coding received signals \cite{lavery2014digital, erkilincc2020pon}. Meanwhile, the MIMO equalizer at the OLT can be simplified to a multiple-input-single-output equalizer due to the transmitted single-polarization real-valued signal.

\begin{figure}[!t]
\centering
\includegraphics[width=\linewidth]{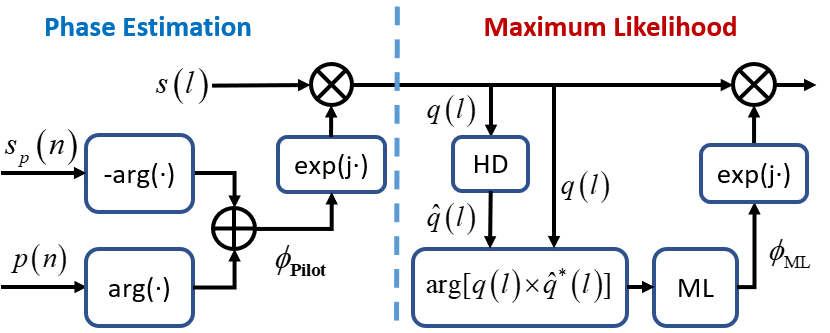}
\caption{The structure of a pilot-based CPE integrated with the maximum likelihood (ML) algorithm.}
\label{CPE}
\end{figure}

\subsection{Pilot-Based CPE}
As Fig. \ref{CPE} shows, a pilot-based CPE integrated with the maximum likelihood algorithm is used to estimate the carrier phase noise with low computational complexity. One pilot symbol is periodically inserted into every $M$ payload symbol for the pilot-based CPE, which estimates the phase noise of the pilot symbol as
\begin{equation}
\phi_{\text {Pilot}}(n)=\arg \left[r_p^{*}(n) \times p(n)\right]
\end{equation}
where $r_p(j)$ is the $n$-th received pilot symbol after the MIMO equalizer and $p(n)$ denotes the $n$-th pilot symbol. The phase noise of the $M-1$ symbols between the $n$-th and the $(n+1)$-th pilot symbols is initialized as $\phi_{\text {Pilot}}(n)$. Then, the following maximum likelihood algorithm is employed to estimate the residual phase noise as
\begin{equation}
\phi_{\text {ML}}(i)=\frac{1}{2 Q+1} \sum_{l=i-Q}^{i+Q} \arg \left[q(l) \times\hat{q}^{*}(l)\right]
\end{equation}
where $q(l)$ denotes the received signal after the carrier phase compensation. $\hat{q}(l)$ is its decision. $Q$ is the half-length of the average filter. After the pilot-based CPE and compensation, the QAM can be decided to regenerate the bit sequence, which can finally be sent into FEC to correct the error bits.

\begin{figure}[!t]
\centering
\includegraphics[width=\linewidth]{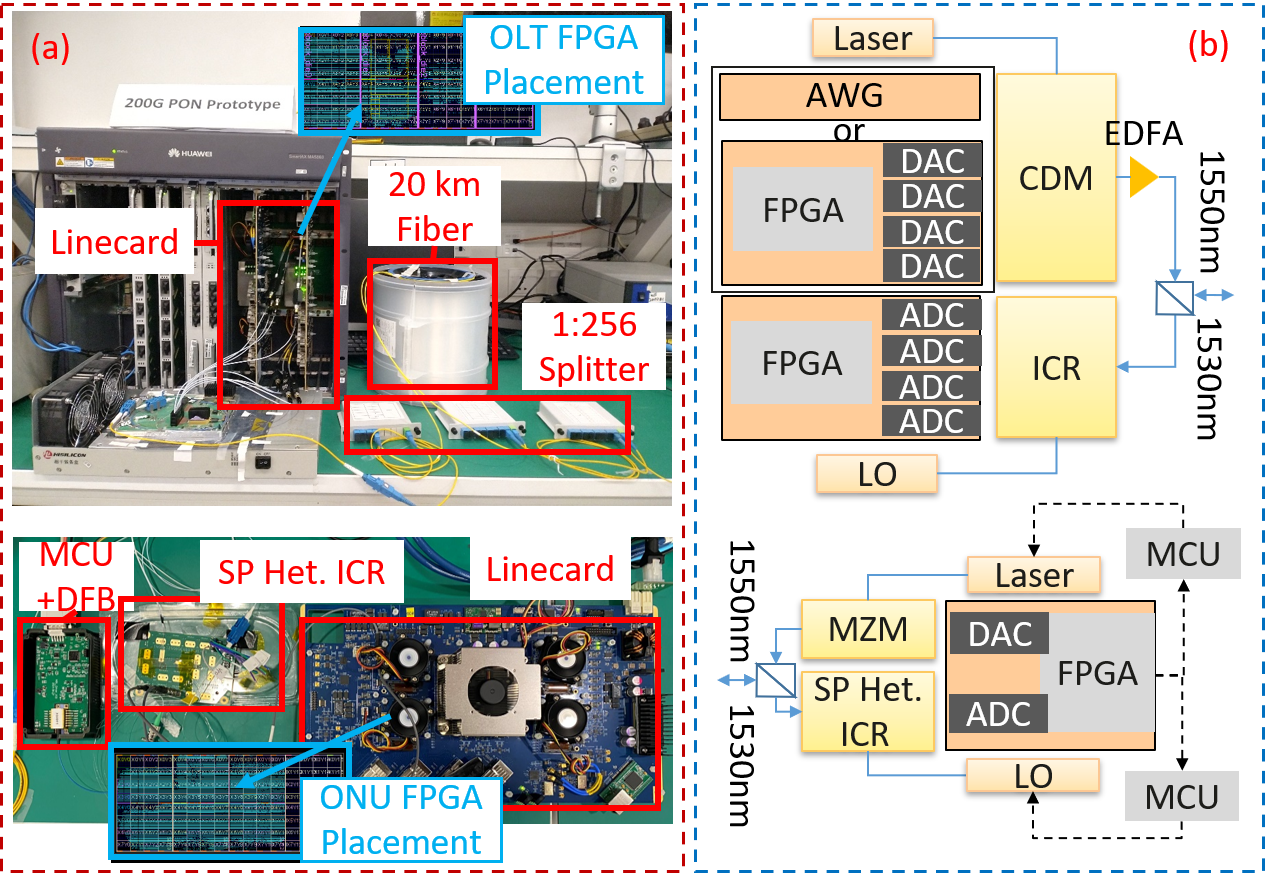}
\caption{(a) The linecards including field-programmable gate array (FPGA) integrated with 31GSa/s DACs and ADCs for upstream and downstream scenarios. (b) Schematic diagrams of the full coherent transceiver and the simplified coherent transceiver. AWG: arbitrary waveform generator. CDM: coherent driver modulator. ICR: integrated coherent receiver. SP: single polarization. Het.: heterodyne.}
\label{RealSetup}
\end{figure}

\section{Experimental Setups and Results} \label{SectionIV}
As Fig. \ref{RealSetup} (a) and Fig. \ref{RealSetup} (b) show, we developed the linecards including field-programmable gate array (FPGA) integrated with 31GSa/s DACs and ADCs for upstream and downstream scenarios. At the OLT, a commercial coherent driver modulator (CDM) and integrated coherent receiver (ICR) were used to implement the full coherent transceiver. One FPGA integrated with four DACs and One FPGA integrated with four ADCs were used to generate the transmitted signal and process the received signal, respectively. At the ONU, single-polarization MZM and single-polarization heterodyne ICR were used to implement the ultra-simple coherent transceiver. Only one FPGA integrated with one DAC and one ADC was employed to generate the transmitted signal and process the received signal. The proposed ultra-simple coherent transceiver for ONU has low cost and low power consumption.

\begin{figure}[!t]
\centering
\includegraphics[width=\linewidth]{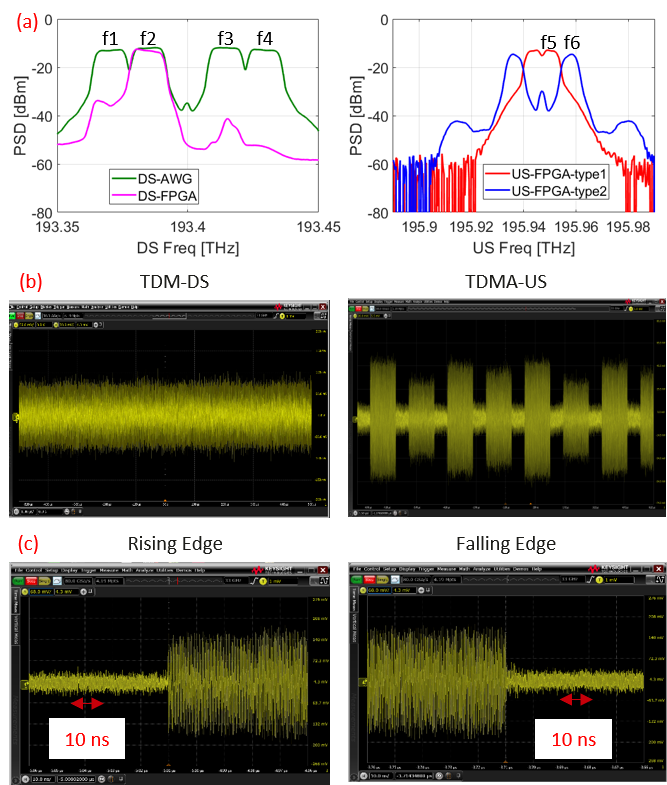}
\caption{(a) Measured spectra of downstream and upstream scenarios. (b) Measured waveforms of TDM downstream and TDMA upstream signals. (c) Rising and falling edges of the burst-mode signals.}
\label{SpectraWave}
\end{figure}

\begin{figure}[!t]
\centering
\includegraphics[width=\linewidth]{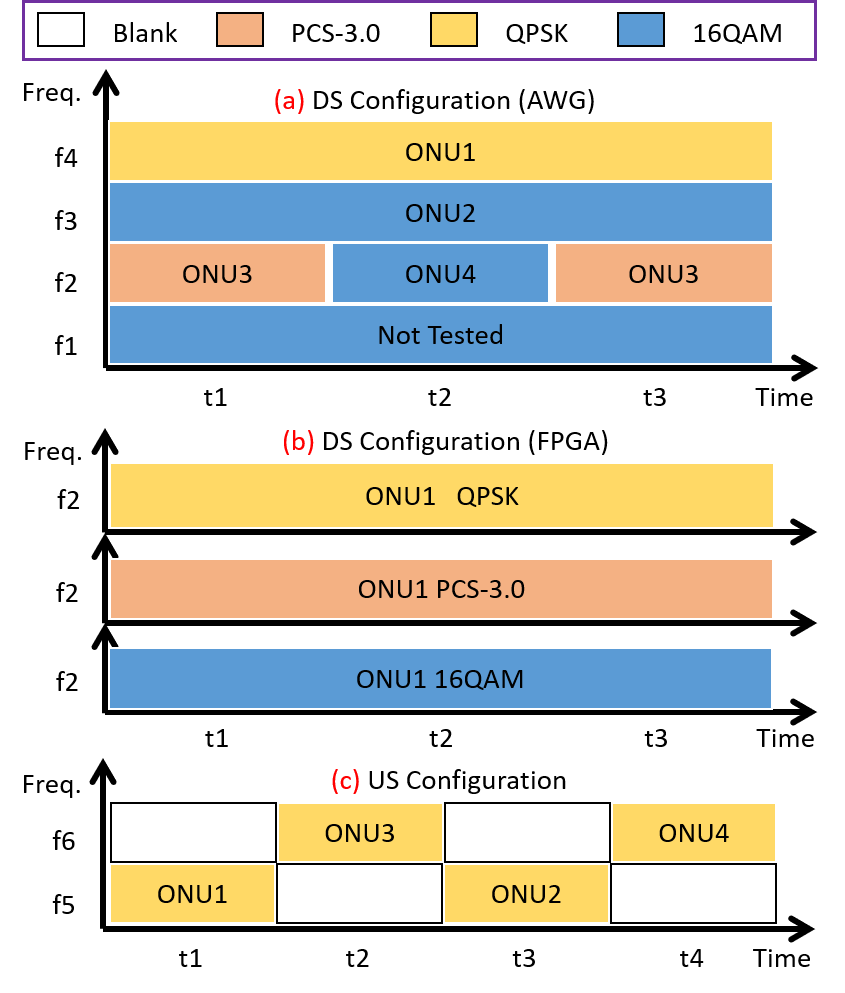}
\caption{Time and frequency source allocations for (a) downstream test with AWG at the transmitter and FPGA at the receiver, (b) downstream test with FPGA, and (c) upstream test with FPGA.}
\label{SourceAlloc}
\end{figure}

\begin{figure}[!t]
\centering
\includegraphics[width=\linewidth]{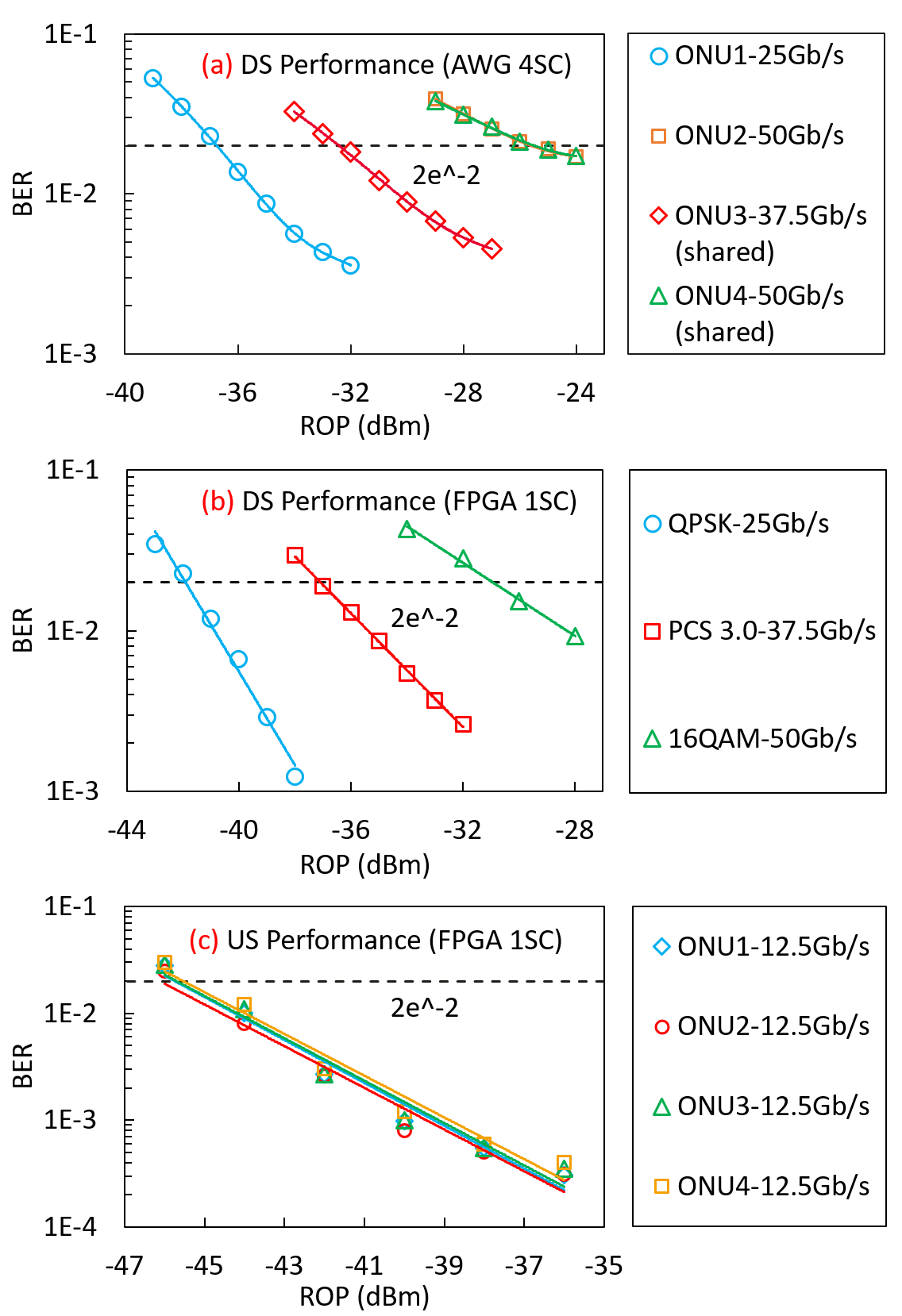}
\caption{The BER versus ROP for (a) downstream test with AWG at the transmitter and FPGA at the receiver, (b) downstream test with FPGA at both transmitter and receiver, and (c) upstream test with FPGA at both transmitter and receiver.}
\label{BERPerf}
\end{figure}

The measured downstream and upstream spectra are shown in Fig. \ref{SpectraWave} (a). Due to the limited sampling rate of the DACs, two experimental cases were done for the downstream scenario: 1) using a linecard to generate one subcarrier with data, and 2) using an AWG to generate four subcarriers with data and one blank subcarrier. In both cases, we used a linecard for implementing the real-time DSP to process the downstream signals at the ONU, and the same card was also used to generate an upstream CAP signal with two subcarriers. Specifically, we generated a CAP signal with either the inner subcarrier or the outer subcarrier at each ONU, and then we used another linecard for the burst-mode DSP to process the upstream signal at the OLT. Fig. \ref{SpectraWave} (b) depicts the waveforms of the downstream TDM signals and the upstream TDMA signals. It is worth noting that the downstream TDM signal is a continuous signal and the upstream TDMA signal is a burst signal. Fig. \ref{SpectraWave} (c) zooms in on the rising edges and the falling edges of the burst signals. The double-sided arrows denote the dimension of 10ns. Obviously, the rising and falling edges are much less than 1ns. Simultaneously, the receiver DSPs can provide real-time estimations of laser frequency offsets. The estimated laser frequency offsets were sent to the microcontroller units (MCUs), which then control the temperature and the driving injection currents of the DFB lasers at the ONUs to achieve real-time frequency locking. 

Figure \ref{SourceAlloc} shows the time and frequency source allocations for the downstream and upstream tests. As Fig. \ref{SourceAlloc} (a) shows, when the AWG was used at the transmitter for the downstream scenario, ONU1 and ONU2 were assigned with the dedicated subcarriers on the frequency of f4 and f3, respectively. ONU3 and ONU4 shared the subcarrier on the frequency of f2, which occupied half of the time slots. Note that although an ONU only detects one subcarrier, it can receive the power of all four data subcarriers. As Fig. \ref{SourceAlloc} (b) depicts, when the FPGA linecard was used at the transmitter of OLT, we only tested the dedicated subcarrier cases with different spectral efficiency. Fig. \ref{SourceAlloc} (c) depicts that ONU1 and ONU2 were assigned to different time slots of the subcarrier on the frequency of f5 for the upstream scenario. ONU3 and ONU4 shared the subcarrier on the frequency of f6.

As Fig. \ref{BERPerf} (a) depicts, to achieve total line rates of 100Gb/s, 150Gb/s, and 200Gb/s in the downstream scenario (assuming sending the same modulation formats to all ONUs), the required ROPs are $-36.8$dBm, $-32$dBm, and $-25$dBm at the BER threshold of $2\times 10^{-2}$, respectively. $15 \%$ overhead soft-decision FEC with a BER threshold of $2\times 10^{-2}$ is considered. An EDFA was employed after the CDM at the OLT to amplify optical output power to 7dBm. Therefore, the optical power budgets are 43.8dB, 39dB, and 32dB for line rates of 100Gb/s, 150Gb/s, and 200Gb/s, respectively. It is worth noting that the EDFA can be replaced by a cost-effective booster SOA for commercial deployment. As Fig. \ref{BERPerf} (b) shows, the required ROPs for 25Gb/s QPSK, 37.5Gb/s PCS-16QAM, and 50Gb/s 16QAM flexible-rate transmission are $-41.8$dBm, $-37$dBm, and $-31$dBm, respectively. The power of one subcarrier was measured as the ROP in the test cases.

Switching from O-band wavelength to C-band wavelength can reduce fiber loss. The measured total loss of 1:256 splitter and 20 km fiber is 31 dB in the lab. The situation in a real-world network may be worse, and it might be challenging to achieve 1:256 splitting ratio when 16QAM is modulated. Fortunately, the optical power budgets of PCS-16QAM and QPSK are less stringent. With the development of high-bandwidth transceiver devices, it is possible to further increase the symbol rate and reduce the entropy of PCS-16QAM. Thus, it is promising to simultaneously achieve 1:256 splitting ratio and 200 Gb/s downstream line rate in the real-world network.

Figure \ref{BERPerf} (c) shows that the required ROP is approximately $-45$dBm for the upstream scenario with 12.5Gb/s per ONU. The ONUs have the ability to transmit data on both the inner and the outer CAP with two subcarriers for achieving the maximum rate of 25Gb/s per ONU. The required ROP should be approximately $-42$dBm for the upstream scenario with 25Gb/s per ONU. We used one single MZM at the transmitter of each ONU, thus the insertion loss of the ONU transmitter is much less than that of the OLT transmitter. The launch optical power was approximately 1dBm without any optical amplifier at the ONUs. Therefore, the optical power budget of the upstream scenario can achieve 43dB for a peak line rate of 100Gb/s.

\begin{figure}[!t]
\centering
\includegraphics[width=0.95\linewidth]{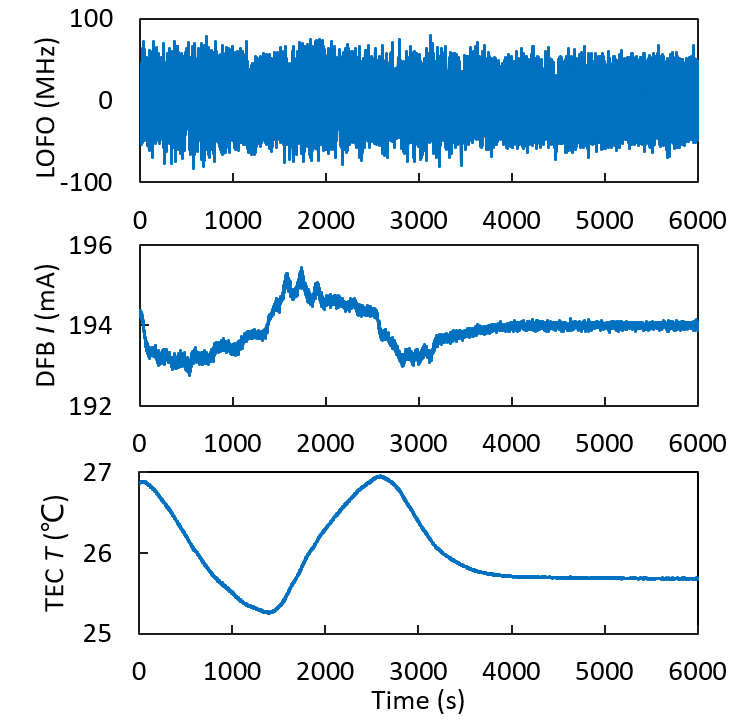}
\caption{The real-time frequency-locking results for the DFB laser when the environment temperature was varied between 10 $^\text{o}$C and 60$^\text{o}$C.}
\label{Laser}
\end{figure}

Fig. \ref{Laser} shows the real-time frequency-locking results for the DFB laser when the environment temperature was varied between 10$^\text{o}$C and 60$^\text{o}$C. The topic subfigure shows the estimated frequency offset of the DFB laser using the DSP. The middle and bottom subfigures show the injection current and target temperature of TEC (Thermo-Electric Cooler) set by the MCU, respectively. When the environment temperature was changed, the injection current and target temperature of TEC transitorily fluctuated and immediately settled down. By controlling both the current and the temperature, the frequency offset is strictly confined within ($-100$, $100$) MHz, which can be well compensated by the receiver DSP. The experimental results verify the cost-effective DFB laser can meet the requirement for the coherent PON.

\section{Conclusions} \label{SectionV}
In this paper, we introduce the architectures, algorithms, and demonstrations for the TFDMA-based coherent PONs. The system architectures consist of a fully coherent transceiver at the OLT and an ultra-simple coherent transceiver at the ONU, which can greatly reduce the cost and power consumption of the ONU. Meanwhile, the Almouti coding and specific spectra are designed to implement the proposed system architectures. Fast and low-complexity DSP algorithms are used for processing upstream and downstream signals. Based on the system architectures and the DSP algorithms, the first real-time TFDMA-based coherent PON was experimentally demonstrated to support at most 256 end users, and the peak line rates of 100Gb/s and 200Gb/s in the upstream and downstream scenarios, respectively. Meanwhile, the real-time frequency locking within a frequency offset of ($-100$, $100$) MHz verifies the feasibility of a cost-effective DFB laser. In conclusion, the system architectures, DSP algorithms, and real-time demonstrations enable TFDMA-based coherent PON in the future beyond 50G PON.

\bibliographystyle{IEEEtran}
\bibliography{bibtex/bib/IEEEexample}

\begin{thebibliography}{10}
\providecommand{\url}[1]{#1}
\csname url@samestyle\endcsname
\providecommand{\newblock}{\relax}
\providecommand{\bibinfo}[2]{#2}
\providecommand{\BIBentrySTDinterwordspacing}{\spaceskip=0pt\relax}
\providecommand{\BIBentryALTinterwordstretchfactor}{4}
\providecommand{\BIBentryALTinterwordspacing}{\spaceskip=\fontdimen2\font plus
\BIBentryALTinterwordstretchfactor\fontdimen3\font minus
  \fontdimen4\font\relax}
\providecommand{\BIBforeignlanguage}[2]{{%
\expandafter\ifx\csname l@#1\endcsname\relax
\typeout{** WARNING: IEEEtran.bst: No hyphenation pattern has been}%
\typeout{** loaded for the language `#1'. Using the pattern for}%
\typeout{** the default language instead.}%
\else
\language=\csname l@#1\endcsname
\fi
#2}}
\providecommand{\BIBdecl}{\relax}
\BIBdecl

\bibitem{ITUT202150gigabit}
``{50-Gigabit-capable passive optical networks (50G-PON): Physical media
  dependent (PMD) layer specification},'' ITU-T G.9804.3, 2021.

\bibitem{9123509}
D.~Zhang, D.~Liu, X.~Wu, and D.~Nesset, ``{Progress of ITU-T higher speed
  passive optical network (50G-PON) standardization},'' \emph{Journal of
  Optical Communications and Networking}, vol.~12, no.~10, pp. D99--D108, 2020.

\bibitem{ITUT2023b50gigabit}
``{PON transmission technologies above 50 Gb/s per wavelength},'' ITU-T work
  programme G. suppl. VHSP, 2023.

\bibitem{zhang2022coherent}
J.~Zhang and Z.~Jia, ``{Coherent Passive Optical Networks for
  100G/$\lambda$-and-Beyond Fiber Access: Recent Progress and Outlook},''
  \emph{IEEE Network}, vol.~36, no.~2, pp. 116--123, 2022.

\bibitem{suzuki2022digital}
N.~Suzuki, H.~Miura, K.~Mochizuki, and K.~Matsuda, ``{Digital Coherent based
  PON Technologies and Beyond-100G Optical Access Systems},'' in \emph{2022
  27th OptoElectronics and Communications Conference (OECC) and 2022
  International Conference on Photonics in Switching and Computing
  (PSC)}.\hskip 1em plus 0.5em minus 0.4em\relax IEEE, 2022, pp. 1--3.

\bibitem{teixeira2016coherent}
A.~Teixeira, A.~Shahpari, R.~Ferreira, F.~P. Guiomar, and J.~D. Reis,
  ``{Coherent access},'' in \emph{Optical Fiber Communication
  Conference}.\hskip 1em plus 0.5em minus 0.4em\relax Optica Publishing Group,
  2016, pp. M3C--5.

\bibitem{lavery2018recent}
D.~Lavery, S.~Erk{\i}l{\i}n{\c{c}}, P.~Bayvel, and R.~I. Killey, ``{Recent
  Progress and Outlook for Coherent PON},'' in \emph{Optical Fiber
  Communication Conference}.\hskip 1em plus 0.5em minus 0.4em\relax Optica
  Publishing Group, 2018, pp. M3B--1.

\bibitem{faruk2020coherent}
M.~S. Faruk and S.~J. Savory, ``{Coherent access: Status and opportunities},''
  in \emph{2020 IEEE Photonics Society Summer Topicals Meeting Series
  (SUM)}.\hskip 1em plus 0.5em minus 0.4em\relax IEEE, 2020, pp. 1--2.

\bibitem{zhang2021efficient}
J.~Zhang, Z.~Jia, M.~Xu, H.~Zhang, and L.~A. Campos, ``{Efficient preamble
  design and digital signal processing in upstream burst-mode detection of 100G
  TDM coherent-PON},'' \emph{Journal of Optical Communications and Networking},
  vol.~13, no.~2, pp. A135--A143, 2021.

\bibitem{xu2022intelligent}
M.~Xu, Z.~Jia, H.~Zhang, L.~A. Campos, and C.~Knittle, ``{Intelligent Burst
  Receiving Control in 100G Coherent PON with 4$\times$ 25G TFDM Upstream
  Transmission},'' in \emph{Optical Fiber Communication Conference}.\hskip 1em
  plus 0.5em minus 0.4em\relax Optica Publishing Group, 2022, pp. Th3E--2.

\bibitem{zhang2023low}
H.~Zhang, Z.~Jia, L.~A. Campos, and C.~Knittle, ``{Low-Cost 100G Coherent PON
  Enabled by TFDM Digital Subchannels and Optical Injection Locking},'' in
  \emph{Optical Fiber Communication Conference}.\hskip 1em plus 0.5em minus
  0.4em\relax Optica Publishing Group, 2023, pp. W1I--4.

\bibitem{welch2022digital}
D.~Welch, A.~Napoli, J.~B{\"a}ck, N.~Swenson, W.~Sande, J.~Pedro, F.~Masoud,
  A.~Chase, C.~Fludger, H.~Sun \emph{et~al.}, ``{Digital Subcarriers: A
  Universal Technology for Next Generation Optical Networks},'' in
  \emph{Optical Fiber Communication Conference}.\hskip 1em plus 0.5em minus
  0.4em\relax Optica Publishing Group, 2022, pp. Tu3H--1.

\bibitem{welch2022digital-1}
D.~Welch, A.~Napoli, J.~B{\"a}ck, S.~Buggaveeti, C.~Castro, A.~Chase, X.~Chen,
  V.~Dominic, T.~Duthel, T.~A. Eriksson \emph{et~al.}, ``{Digital subcarrier
  multiplexing: enabling software-configurable optical networks},''
  \emph{Journal of Lightwave Technology}, vol.~41, no.~4, pp. 1175--1191, 2022.

\bibitem{hosseini2022multi}
M.~M. Hosseini, J.~Pedro, N.~Costa, A.~Napoli, J.~E. Prilepsky, and S.~K.
  Turitsyn, ``{Multi-period Planning in Metro-Aggregation Networks using
  Point-to-Multipoint Transceivers},'' in \emph{IEEE Global Communications
  Conference}.\hskip 1em plus 0.5em minus 0.4em\relax IEEE, 2022, pp.
  2921--2926.

\bibitem{xing2022first}
S.~Xing, G.~Li, J.~Chen, J.~Zhang, N.~Chi, Z.~He, and S.~Yu, ``{First
  Demonstration of PS-QAM based Flexible Coherent PON in Burst-Mode with 300G
  Peak-Rate and Record Dynamic-Range and Net-Rate Product up to 7,104
  dB{\textperiodcentered} Gbps},'' in \emph{Optical Fiber Communications
  Conference}.\hskip 1em plus 0.5em minus 0.4em\relax IEEE, 2022, p. Th4A.4.

\bibitem{wei2022time}
Z.~Wei, J.~Zhang, W.~Li, and D.~V. Plant, ``{Time-variant entropy regulated
  multiple access for flexible coherent PON},'' \emph{Optics Letters}, vol.~47,
  no.~19, pp. 5148--5151, 2022.

\bibitem{zhang2023flexible}
J.~Zhang, G.~Li, S.~Xing, and N.~Chi, ``{Flexible and adaptive coherent PON for
  next-generation optical access network},'' \emph{Optical Fiber Technology},
  vol.~75, p. 103190, 2023.

\bibitem{faruk2022experimental}
M.~S. Faruk, X.~Li, and S.~J. Savory, ``{Experimental demonstration of
  100/200-Gb/s/$\lambda$ PON downstream transmission using simplified coherent
  receivers},'' in \emph{Optical Fiber Communication Conference}.\hskip 1em
  plus 0.5em minus 0.4em\relax Optica Publishing Group, 2022, pp. Th3E--5.

\bibitem{hraghi2022analysis}
A.~Hraghi, G.~Rizzelli, A.~Pagano, V.~Ferrero, and R.~Gaudino, ``{Analysis and
  experiments on C band 200G coherent PON based on Alamouti
  polarization-insensitive receivers},'' \emph{Optics Express}, vol.~30,
  no.~26, pp. 46\,782--46\,797, 2022.

\bibitem{li2022bidirectional}
X.~Li, M.~S. Faruk, and S.~J. Savory, ``{Bidirectional symmetrical 100
  Gb/s/$\lambda$ coherent PON using a simplified ONU transceiver},'' \emph{IEEE
  Photonics Technology Letters}, vol.~34, no.~16, pp. 838--841, 2022.

\bibitem{gaudino2012use}
R.~Gaudino, V.~Curri, G.~Bosco, G.~Rizzelli, A.~Nespola, D.~Zeolla,
  S.~Straullu, S.~Capriata, and P.~Solina, ``{On the use of DFB lasers for
  coherent PON},'' in \emph{Optical Fiber Communication Conference}.\hskip 1em
  plus 0.5em minus 0.4em\relax Optica Publishing Group, 2012, pp. OTh4G--1.

\bibitem{Xing:23}
Z.~Xing, K.~Zhang, X.~Chen, Q.~Feng, K.~Zheng, Y.~Zhao, Z.~Dong, J.~Zhou,
  T.~Gui, Z.~Ye, and L.~Li, ``{First Real-time Demonstration of 200G TFDMA
  Coherent PON using Ultra-simple ONUs},'' in \emph{Optical Fiber Communication
  Conference}.\hskip 1em plus 0.5em minus 0.4em\relax Optica Publishing Group,
  2023, p. Th4C.4.

\bibitem{liu2023flexible}
G.~Liu, J.~Zhou, Y.~Huang, G.~Wang, Y.~Bo, Y.~Wu, Y.~Lu, J.~He, M.~Li, Z.~Ye
  \emph{et~al.}, ``{Flexible transceiver for an access network: a multicarrier
  entropy loading approach},'' \emph{Journal of Optical Communications and
  Networking}, vol.~15, no.~7, pp. 442--448, 2023.

\bibitem{zhou2022100g}
J.~Zhou, J.~He, X.~Lu, G.~Wang, Y.~Bo, G.~Liu, Y.~Huang, L.~Li, C.~Yang,
  H.~Wang \emph{et~al.}, ``{100G fine-granularity flexible-rate passive optical
  networks based on discrete multi-tone with PAPR optimization},''
  \emph{Journal of Optical Communications and Networking}, vol.~14, no.~11, pp.
  944--950, 2022.

\bibitem{shen2023demonstration}
W.~Shen, S.~Xing, G.~Li, Z.~Li, A.~Yan, J.~Wang, J.~Zhang, and N.~Chi,
  ``{Demonstration of Beyond 100G Three-Dimensional Flexible Coherent PON in
  Downstream with Time, Frequency and Power Resource Allocation Capability},''
  in \emph{Optical Fiber Communication Conference}.\hskip 1em plus 0.5em minus
  0.4em\relax Optica Publishing Group, 2023, pp. W1I--5.

\bibitem{wang2023fast}
H.~Wang, J.~Zhou, Z.~Xing, Q.~Feng, K.~Zhang, K.~Zheng, X.~Chen, T.~Gui, L.~Li,
  J.~Zeng, J.~Yang, W.~Liu, C.~Yu, and Z.~Li, ``{Fast-Convergence Digital
  Signal Processing for Coherent PON Using Digital SCM},'' \emph{Journal of
  Lightwave Technology}, vol.~41, no.~14, pp. 4635--4643, 2023.

\bibitem{savory2010digital}
S.~J. Savory, ``{Digital coherent optical receivers: Algorithms and
  subsystems},'' \emph{IEEE Journal of selected topics in quantum electronics},
  vol.~16, no.~5, pp. 1164--1179, 2010.

\bibitem{zhang2020efficient}
J.~Zhang, Z.~Jia, M.~Xu, H.~Zhang, and L.~A. Campos, ``{Efficient preamble
  design and digital signal processing in upstream burst-mode detection of 100G
  TDM coherent-PON},'' \emph{Journal of Optical Communications and Networking},
  vol.~13, no.~2, pp. A135--A143, 2020.

\bibitem{morelli1999improved}
M.~Morelli and U.~Mengali, ``{An improved frequency offset estimator for OFDM
  applications},'' in \emph{1999 IEEE Communications Theory Mini-Conference
  (Cat. No. 99EX352)}.\hskip 1em plus 0.5em minus 0.4em\relax IEEE, 1999, pp.
  106--109.

\bibitem{pittala2014training}
F.~Pittala, I.~Slim, A.~Mezghani, and J.~A. Nossek, ``{Training-aided
  frequency-domain channel estimation and equalization for single-carrier
  coherent optical transmission systems},'' \emph{Journal of Lightwave
  Technology}, vol.~32, no.~24, pp. 4849--4863, 2014.

\bibitem{lavery2014digital}
D.~Lavery and S.~J. Savory, ``{Digital Coherent Technology for Long-Reach
  Optical Access},'' in \emph{Optical Fiber Communication Conference}.\hskip
  1em plus 0.5em minus 0.4em\relax Optica Publishing Group, 2014, pp. Tu2F--1.

\bibitem{erkilincc2020pon}
M.~Erk{\i}l{\i}n{\c{c}}, R.~Emmerich, K.~Habel, V.~Jungnickel,
  C.~Schmidt-Langhorst, C.~Schubert, and R.~Freund, ``{PON transceiver
  technologies for $\geq$ 50 Gbits/s per $\lambda$: Alamouti coding and
  heterodyne detection},'' \emph{Journal of Optical Communications and
  Networking}, vol.~12, no.~2, pp. A162--A170, 2020.

\end{thebibliography}

\end{document}